# Electronic structure and thermoelectric properties of epitaxial Sc$_{1-x}$V$_x$N$_y$ thin films grown on MgO(001)

Susmita Chowdhury[*], Niraj Kumar Singh, Sanath Kumar Honnali, Grzegorz Greczynski, Per Eklund, Arnaud le Febvrier, and Martin Magnuson

*Thin Film Physics Division, Department of Physics, Chemistry and Biology (IFM), Linköping University, Linköping SE-581 83, Sweden*

## Abstract

The electronic structure of Sc$_{1-x}$V$_x$N$_y$ epitaxial films with different alloying concentrations of V are investigated with respect to effects on thermoelectric properties. Band structure calculations on Sc$_{0.75}$V$_{0.25}$N indicate that V *3d* states lie in the band gap of the parent ScN compound in the vicinity of the Fermi level. Thus, theoretically the presence of light (dispersive) bands at the Γ-point with band multiplicity is expected to lead to lower electrical resistivity while flat (heavy) bands at X-W-K symmetry points are associated with higher Seebeck coefficient than that of ScN. With this aim, epitaxial Sc$_{1-x}$V$_x$N$_y$ thin film samples were deposited on MgO(001) substrates. All the samples showed N substoichiometry and pseudocubic crystal structure. The N-vacancy-induced states were visible in the Sc *2p* XAS spectra. The reference ScN and Sc$_{1-x}$V$_x$N$_y$ samples up to $x = 0.12$ were *n* type, exhibiting carrier concentration of $10^{21}$ cm$^{-3}$, typical for degenerate semiconductors. For the highest V alloying of $x = 0.15$, holes became the majority charge carrier as indicated by the positive Seebeck coefficient. The underlying electronic structure and bonding mechanism in Sc$_{1-x}$V$_x$N$_y$ influence the electrical resistivity, Seebeck coefficient, and Hall effect. Thus, the work contributes to the fundamental understanding of the correlated defects and thermoelectric properties to the electronic structure in Sc-N system with V alloying.

## 1. Introduction

Conventional thermoelectric materials such as tellurides have been used since long to harvest power from waste heat [1–4]. These materials have limitations because of toxicity and scarcity [5,6]. There is thus a great deal of development of nontraditional thermoelectric materials, *e.g.*, Heusler alloys, Zintl phases, silicides, oxides, and selenides [4,7–9]. Recently, the semiconducting early transition metal nitrides based on ScN and CrN have emerged as alternative thermoelectric materials and are noteworthy compared to the materials listed above because of abundance, cost-effectiveness, and high symmetry rocksalt crystal structure. The simple cubic structure provides pathways for low electrical resistivity (ρ) with high valley degeneracy and low thermal conductivity by phonon anharmonicity [10].

The efficiency of any thermoelectric material is determined by the dimensionless thermoelectric figure-of-merit expressed as, $zT = (S^2\sigma/\kappa_e+\kappa_l)T$. Here, S = Seebeck coefficient, σ = electrical conductivity, $\kappa_e$ and $\kappa_l$ = electronic and lattice thermal conductivity, respectively, and T = absolute temperature. Thus, *zT* can be optimized by simultaneous increase of $S^2\sigma$ (known as the power factor) and lowering of κ (= $\kappa_e+\kappa_l$). However, in practice this is difficult to achieve as S is inversely coupled to both σ and $\kappa_e$ [11]. Typical power factor of ScN is 2.5-3.5×10$^{-3}$ W m$^{-1}$ $K^{-2}$) at 600-840 K [11,12] and CrN of 1.5-5×$10^{-3}$ W $m^{-1}$ $K^{-2}$ at 300 K [13,14], comparable to $Bi_2Te_3$ at 350 K and PbTe at 500 K [15,16]. The thermal conductivity of CrN-based thin films is ~4 W $m^{-1}$ $K^{-1}$ [13], but the thermal conductivity of ScN is typically 10-12 W $m^{-1}$ $K^{-1}$ [17] at room temperature and requires reduction [11]. Since natural $^{45}$Sc has no isotopes, mass contrast due to isotope reduction has no effect on the thermal conductivity of pure ScN.

Hence, defects induced by doping/alloying (Cr, Nb) [18,19] and domain size reduction [17] in ScN have been used as a mechanism to reduce the thermal conductivity of ScN by increasing the phonon scattering. Furthermore, vacancies (≤3 atomic %) in the Sc (p-type) or N (n-type) site can retain high $S$ with low metal-like $\rho$ increasing the overall power factor [20]. Alloying and ion-implantation (*i.e.*, doping) with the same element (*e.g.* Mg) in ScN alters two different properties. The former enhances the power factor while the latter reduces the thermal conductivity in ScN [21,22]. Defects induced by ion implantations ($Li^+$, $Ar^+$ and $He^+$) in ScN increases $\rho$, but results in an increase in $S$ and reduction in κ by overall increasing the $zT$ [23–25].

The choice of suitable alloying elements for possible enhancement of the power factor is reflected in the electronic density of states (DOS) near the Fermi level ($E_F$) or by modulating the band structure [26]. According to the Mott equation for degenerate semiconductors and metals [27], $S$ can be expressed as [2],

$$S = -\frac{\pi^2}{3}\frac{k_B^2 T}{q}\left[\frac{d\{\ln\sigma(E)\}}{dE}\right]_{E=E_F}$$

Here, electrical conductivity $\sigma(E) \propto DOS$, since the assumption is made that the electronic scattering is independent of energy [28]. Thus, a sharp slope with local increase in the *DOS* accompanied by asymmetry in the energy distribution function is key to achieve enhanced thermoelectric properties [20,29].

The electronic structure of ScN has been studied based on the contributions of different hybridization states of Sc and N [30] to understand the role of point defects (N vacancy, O and C substitution of the N-site, Mg, Be and Zr substitution of the Sc-site) on the thermoelectric properties of ScN [20]. Substantial amount of theoretical research using band structure calculations have been dedicated to obtain a correct bandgap of ScN, which is ~0.9 eV [11,31]. Alloying of V in ScN is expected to yield a stable solid solubility with low negative enthalpy of mixing due to small lattice mismatch [32]. The atomic mass of V (50.94 g $mol^{-1}$) is slightly higher than Sc (44.96 g $mol^{-1}$) with two more valence electrons, which is expected to change the carrier concentration ($n$) in stoichiometric ScN. Further, $n$ is related to $S$ ($\propto 1/n^{2/3}$) and $\rho$ ($\propto 1/n$) [11], which should alter the thermoelectric properties. This underscores the need to investigate the electronic band structure of $Sc_{1-x}V_xN_y$ near $E_F$, correlating the electronic structure and thermoelectric properties of $Sc_{1-x}V_xN_y$.

In the present work, we probe the electronic structure and thermoelectric properties with alloying of V in ScN thin films. The structural and electronic properties of the alloyed samples are studied with a combination of experimental and theoretical methods, to elucidate the correlations between the charge carrier concentration, carrier mobility (μ) and thermoelectric properties to the underlying electronic structure and chemical bonding.

## 2. Methods

2.1. Computational details

Electronic structure calculations were performed within a density-functional-theory framework and the projector augmented wave (PAW) method [33] as implemented in the Vienna *ab initio* simulation package (VASP) 6.3.0 [34,35] using a self-consistent energy convergence criteria of $1\times10^{-5}$ eV. Convergence during structural relaxation was obtained by minimizing the forces on the ions to <0.001 eV/Å using a 29×29×29 k-point mesh. GGA [36] with a combination of a Hubbard Coulomb term (GGA+U) [37,38] was used for treating electron exchange-correlation effects. Hubbard terms of U=3.5 eV was applied to the Sc *3d* and V *3d* orbitals. The energy cutoff for plane waves included in the expansion of wave functions was 700 eV. The ScN unit cell contained 4 Sc atoms and 4 N atoms, while the (Sc,V)N cell contained 3 Sc atoms, 1 V atom and 4 N atoms exhibiting cubic rocksalt NaCl B1 type structure. Sampling of the Brillouin zone was done using a Monkhorst-Pack scheme [39] on a grid of 20×20×20 k-points. For calculating the band structure, a *k*-mesh of 80 points per line between the symmetry points was used. In the present study, we avoid the supercell calculations with doping (instead of alloying) of V in ScN due to the undesired band folding across the symmetry points which may lead to misleading results in the theoretical band structure calculations.

2.2. Experimental Details

Epitaxial $Sc_{1-x}V_xN_y$ and reference ScN thin film samples were deposited on single-side polished single crystalline MgO(001) substrates by dc magnetron sputtering [40]. Prior to deposition, the MgO(001) substrates were cleaned following the procedures referred by le Febvrier *et al.* [41] and post in-situ annealed at 800°C for 30 mins. Sc (2-inch, purity 99.9%) and V (2-inch, purity 99.5%) target power was varied from 110-98 W and 0-15 W, respectively depending on the alloying concentration. The base pressure was ≤2.6×10$^{-7}$ Pa (2×10$^{-9}$ Torr) and the substrate temperature during deposition was 800°C. The Ar and $N_2$ flow were kept constant at 13 sccm (30%) and 31 sccm (70%) keeping the total flow always constant at 44 sccm. The substrate rotation was fixed at 15 rpm to achieve better uniformity. More details about the deposition chamber can be found elsewhere [40].

The compositions of the samples were investigated by Energy Dispersive X-ray Spectroscopy (EDS) and time-of-flight Elastic Recoil Detection Analysis (ToF- ERDA). EDS measurements were performed at an accelerating voltage of 20 kV in a Zeiss Sigma 300 Scanning Electron Microscope (SEM) equipped with an X-Max 80 mm$^2$ EDS detector from Oxford Instruments. The data was sampled and analyzed using AZtec software. ERDA measurements were performed using 36 MeV primary iodine ion ($^{127}I^{8+}$) beam at the Pelletron Tandem accelerator (5 MV NEC-5SDH-2), Uppsala University, Sweden [42]. The incident beam angle to the sample surface normal was 67.5°, while the ToF-telescope and the gas ionization detector were placed at 45° relative to the incident beam direction. The depth profile of the elemental compositions were acquired from the ToF-ERDA time and energy coincidence spectra using the *POTKU* 2.0 software [43]. In addition, complementary RBS measurements were also done on the samples and discussed in the supplementary information (section 1 in the SI).

The out-of-plane X-Ray Diffraction (XRD) measurements were performed using a Malvern PANalytical X'Pert PRO MPD diffractometer equipped with a Cu-Kα source in Bragg-Brentano (θ-2θ) geometry. The operating voltage and current were set at 45 kV and 40 mA. The incident optics consisted of a 1/2° divergence slit and 1/2° anti-scatter slit. The diffracted optics were composed of a 5.0 mm anti-scatter slit, a 0.04 rad Soller slit, a Ni-filter, and an X'Celerator detector. Subsequently, the Bragg-Brentano θ-2θ scans at chi, χ = 0° and 45° were performed using a Malvern PANalytical X'Pert MRD diffractometer in line mode using a hybrid mirror with a divergence slit of 1/2° in the incident beam optics. At the diffracted beam optics, a triple axis monochromator compatible to a hybrid mirror equipped with receiving slits of 1/8° (at χ = 0°) or 1° (for χ = 45°) were used. The ϕ-scans and X-ray reflectivity (XRR) (discussed in section 2 of the SI) measurements were also performed using a Malvern PANalytical X'Pert MRD diffractometer with a similar optics and 1° receiving slit. The voltage and current were set at 45 kV and 40 mA, respectively.

The soft X-ray absorption near-edge spectroscopy (XANES) and resonant inelastic X-ray scattering (RIXS) data [41,42] were measured at the SPECIES beamline, MAX IV, Lund, Sweden at the Sc *2p*-edge, V *2p*-edge, and N *1s*-edge. The XANES spectra were recorded in both total electron yield (surface-sensitive) and total fluorescence yield (bulk-sensitive) modes. To probe the anisotropic nature (if any) around the electronic structure, the linearly polarized X-rays were incident at a grazing angle of 20° either in-plane (a-b plane) to the substrate (vertical polarization) or nearly 80% along the c-axis (horizontal polarization). The energy resolutions were 0.30 eV and 0.35 eV for Sc, and V *2p* absorption edges, respectively. For normalization of the absorption edges, a 4 μm thick Au reference foil was measured directly after each scan. For RIXS, the total energy resolution for the Sc *L* – N *K* emission was 0.58 eV and 0.62 eV for the V *L*-emission. The base pressure was ≤$6.7\times10^{-7}$ Pa during all the measurements. To minimize self-absorption effects [46], the angle of incidence was 20° from the surface plane during the emission measurements. The X-ray photons were detected parallel to the polarization vector of the incoming beam to minimize elastic (Rayleigh) scattering.

Core-level X-ray photoelectron spectroscopy (XPS) (discussed in section 3 of the SI) and the valence band spectroscopy (VBS) were performed on freshly deposited samples. An Axis Ultra DLD instrument, Kratos Analytical (UK) equipped with monochromatized Al-$K_\alpha$ radiation (1486.6 eV) was used. The base pressure was ~$1.3\times10^{-7}$ Pa during measurements. Prior to measurements, samples were sputter etched for 10 minutes using 0.5 keV Ar$^+$ ions at an incident angle of 70° from the surface normal. Such pretreatment has been shown to minimize the sputter damage effects in transition metal nitrides [47]. The area affected by the Ar$^+$ beam was 3×3 mm$^2$, while the analysis area was 0.3×0.7 mm$^2$ (centered in the middle of the etched crater). The analyzer pass energy was 20 eV, resulting in a full width at half maximum of 0.55 eV for the Ag $3d_{5/2}$ peak from the reference Ag sample. The charge referencing for all the samples was done by setting the low energy (density of states) cut-off to 0 eV. Due to the low electrical conductivity of $Sc_{0.93}V_{0.07}N_{0.87}$, an electron flood gun was used during the data acquisition.

A standard four-point-probe set-up (Jandel Model RM3000) at room temperature was used to measure the sheet resistance of the samples. The electrical resistivity (ρ = sheet resistance multiplied with film thickness) of the samples were calculated. The obtained film thickness was around 130 nm measured using XRR (see Table S1 in the SI). The Seebeck coefficients were measured using a home-built setup [48] which uses a resistive heater placed on one end of the sample providing the desired temperature gradient by controlling the applied voltage via a Keithley 2400 Source measure unit. The charge carrier concentration and mobility were measured by Hall measurements in a home-made setup employing a 0.485 T permanent magnet while measuring the Hall voltage in van der Pauw configuration. All the electrical measurements were performed at room temperature.

## 3. Results

3.1. Band structure calculations of ScN and (Sc,V)N

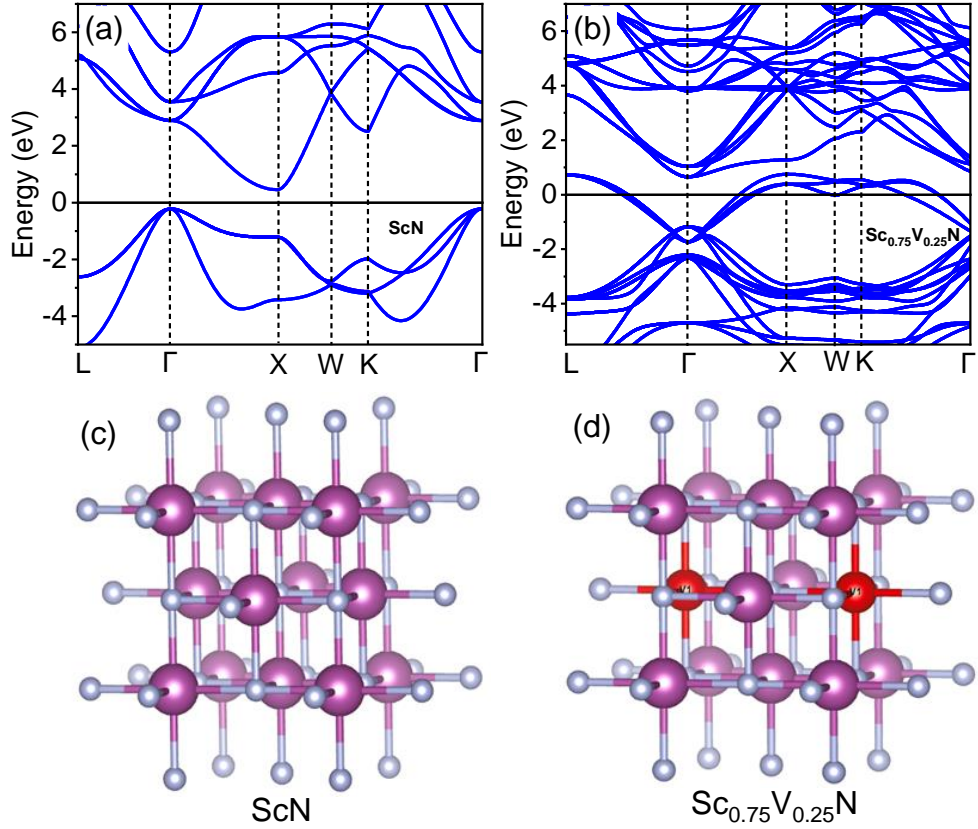

**Figure 1**: Band structures and crystal structures of ScN (a and c) and $Sc_{0.75}V_{0.25}N$ (b and d) in the cubic rocksalt structure at equilibrium volume. The symmetry points are located at Γ(0,0,0), X(0.5,0,0.5), W(0.5,0.25,0.5), L(0.5,0.5,0.5) and K(0.375,0.375,0.375).

Figure 1(a) and (b) show calculated band structures of stoichiometric ScN and $Sc_{0.75}V_{0.25}N$ along the L-Γ-X-W-K-Γ directions in the Brillouin zone. Figure 1(c) and (d) show the cubic rocksalt B1 type crystal structures of ScN and $Sc_{0.75}V_{0.25}N$. For ScN with NaCl B1 structure, the self-consistent calculations resulted in an equilibrium lattice constant (minimum in the total energy) of $a$ = 4.518 Å consistent with literature values [49]. For $Sc_{0.75}V_{0.25}N$ with B1-NaCl structure, the relaxed equilibrium lattice constant was $a$ = 4.410 Å.

For ScN, the conduction band minimum (CBM, above $E_F$) is dominated by Sc $3d$ $t_{2g}$ orbitals while the valence band maximum (VBM) is dominated by electronegative N $2p$ orbitals (below $E_F$) [49]. The VB consists of both heavy (flat) and light (dispersive) hole bands at the Γ-point. Two direct bandgaps of 3.04 eV and 1.64 eV are located at the Γ and X-symmetry points, respectively, and an indirect bandgap of 0.65 eV is located along the Γ-X direction. Initially, the band structure of ScN was debated and assigned to either a semi-metal [50,51] or a semiconductor [52,53], but it is by now long-established that ScN is an indirect (~0.9 eV) narrow bandgap semi-conductor material [54]. Flat (heavy) bands ascribed to high effective electron/hole mass lead to enhanced Seebeck coefficients $S$ [55] that is negative for negatively charged carriers (electrons), and positive for positively charged carriers (holes). In contrast, highly dispersive light bands with low electronic effective mass provide higher charge carrier mobility $\mu$ resulting in an increase in electrical conductivity $\sigma$ ($= ne\mu$) [56]. Thus, the prominent thermoelectric power factor of ScN with metal-like resistivity $\rho$ and high $S$ is reflected in the underlying band structure.

The bands of V $3d$ states occur in the vicinity of $E_F$ with strong V $3d_{xy}$, $3d_{xz}$, $3d_{yz}$ bands crossing the $E_F$ halfway at the $\Gamma$-point. Thus, the $E_F$ is shifted into the CB and an overlap of the CB and VB occurs implying that V alloying drives the system to a metallic-like behavior. This reflects that electrons are the majority charge carriers here indicating $n$ type charge carrier concentrations. The V $3d$ states are dispersive at the $\Gamma$-point but are flat around the X-W-K symmetry points. Thus, the light dispersive bands with band multiplicity (here valley degeneracy) indicates to an increase of $\sigma$ [55–57]. Moreover, the flat band (high effective electron/hole mass) signifies that the *DOS* distribution is localized along the X-W-K direction near to $E_F$ and should result in high $S$. [2,10]

## 3.2. Compositional analysis

Table I shows the elemental composition of the reference ScN and $Sc_{1-x}V_xN_y$ samples obtained from EDS and ToF-ERDA measurements. Within the error bar, the atomic % of metals (Me) match well with each other as obtained from both measurements. The reference ScN is N substoichiometry with a N/Me ratio of 0.87 with ~5% of unintentional O content. With V alloying (4-15%) in ScN, N substoichiometry results in a varying N/Me ratio from 0.81 to 0.87 within the $Sc_{1-x}V_xN_y$ samples. All the $Sc_{1-x}V_xN_y$ samples contain unintentional O which varies between 5.3 - 8.2 at.%. However, (N+O)/Me ratio is nearly equal to 1, indicating that the O possibly occupies the N-site in the crystal structure.

**Table I**. Compositional details of ScN and $Sc_xV_{1-x}N_y$ samples obtained from EDS and ToF-ERDA. Me = metal, and the nomenclature of the samples was given by normalizing the metal site to 100% occupancy.

| Samples | Film composition | | | | | | N/Me ratio | (N+O)/Me ratio |
|---|---|---|---|---|---|---|---|---|
| | EDS (at. % of Me without N) (±1%) | | ERDA (at. %) (±1%) | | | | | |
| | Sc | V | Sc | V | N | O | | |
| $ScN_{0.87}$ | - | - | 50.4 | - | 43.8 | 5.0 | 0.87 | 0.97 |
| $Sc_{0.96}V_{0.04}N_{0.85}$ | 95 | 5 | 48.1 | 2.1 | 42.8 | 5.9 | 0.85 | 0.97 |
| $Sc_{0.93}V_{0.07}N_{0.87}$ | 93 | 7 | 46.7 | 3.4 | 43.6 | 5.3 | 0.87 | 0.97 |
| $Sc_{0.91}V_{0.09}N_{0.83}$ | 89 | 11 | 44.9 | 4.6 | 41.2 | 7.5 | 0.83 | 0.98 |
| $Sc_{0.88}V_{0.12}N_{0.86}$ | 88 | 12 | 43.6 | 6 | 42.6 | 6.8 | 0.86 | 0.99 |
| $Sc_{0.85}V_{0.15}N_{0.81}$ | 83 | 17 | 42.7 | 7.5 | 40.6 | 8.2 | 0.81 | 0.97 |

## 3.3. Structural analysis

### 3.3.1. X-ray diffraction

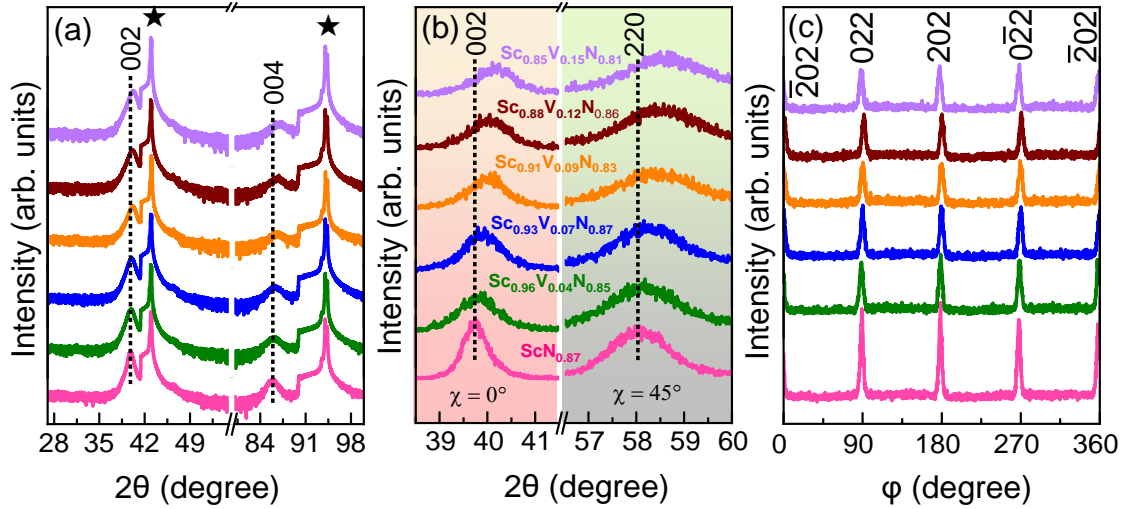

**Figure 2**: (a) 2θ-ω X-ray diffraction patterns of reference $ScN_{0.87}$ and $Sc_{1-x}V_xN_y$ thin film samples deposited on MgO(001) substrates. The stars denote the corresponding substrate peaks. (b) Bragg-Brentano short scan at χ = 0° along [002] and in-plane short scan at χ = 45° along [220] direction. (c) φ-scans performed on (220) plane at χ = 45°.

Figure 2(a) shows out-of-plane XRD data of the $Sc_{1-x}V_xN_y$ and the $ScN_{0.87}$ reference thin film samples deposited on MgO(001) substrates, measured in the Bragg-Brentano geometry. The MgO(001) substrate provides a template for the growth of (002) and higher order (004) reflections for all the samples. Substrate 002 and 004 peaks also appear (indicated by stars at the top of Fig. 2(a)) near the vicinity of the 002 and the 004 peaks of $ScN_{0.87}/Sc_{1-x}V_xN_y$. No secondary phase up to the highest V alloying concentration (up to 15%) is observed indicating formation of a stable solid solution [32].

Figure 2(b) shows out-of-plane and in-plane XRD data keeping the tilt angle χ = 0° and 45°, respectively, in the Bragg-Brentano geometry. In both the 002 and the 220 peaks, a shift along higher diffraction angle is observed upon increasing V content in $Sc_{1-x}V_xN_y$, indicating a decrease of lattice parameters compared to the reference $ScN_{0.87}$ sample. This observation is in accordance with our theoretical calculations along with earlier reported first principles density functional calculations [32]. By fitting the 002 and the 220 peaks by pseudo-Voigt functions, the out-of-plane ($a_\perp$) and in-plane ($a_{//}$) lattice parameters were evaluated and listed in Table II. For the reference $ScN_{0.87}$ sample, the $a_\perp$ and $a_{//}$ values were 4.530 Å and 4.487 Å. For the $Sc_{1-x}V_xN_y$ samples, $a_\perp$ values decrease from 4.522 Å at 4% V content to 4.484 Å at 15% V content, while the $a_{//}$ values change from 4.480 - 4.445 Å (x = 0.04-0.15). The contraction of the cell parameters (lattice parameter/volume) in $Sc_{1-x}V_xN_y$ is caused by the substitution of Sc by the smaller V ion and corresponding changes in the bond strength and bond distance. All the samples have $a_\perp/a_{//}$ ratio in between 1.008 and 1.009 and reveals that the crystal structure of $Sc_{1-x}V_xN_y$ samples is pseudocubic. The slight distortion in the cubic structure deviates from our theoretical crystal structure calculations and previous reports of cubic ScN in the literature [22,54]. We attribute the pseudomorphic distortion of the cubic structure is due to ≤7% of lattice mismatch between the samples and MgO substrate.

Figure 2c shows a φ-scan (at tilt angle χ = 45°, $ScN_{0.87}/Sc_{1-x}V_xN_y$ {220}) of the $Sc_{1-x}V_xN_y$ and the reference $ScN_{0.87}$ samples with different alloying concentrations. Four peaks appear at 90° apart indicating a fourfold symmetry and confirms the epitaxial nature of all the samples. Thus, the epitaxial relationship is $ScN_{0.87}/Sc_{1-x}V_xN_y$ (001)[100] ∥ MgO(001)[100]. The average full width at half maximum (FWHM) of the diffraction peaks for $ScN_{0.87}$ (3.99°) is lower than for the $Sc_{1-x}V_xN_y$ samples where the average FWHM varies from 4.82° to 5.35° (listed in Table II). The increase in the average FWHM for the $Sc_{1-x}V_xN_y$ samples compared to the $ScN_{0.87}$ sample indicates a slightly degraded crystalline quality due to local distortions induced by V alloying in ScN.

**Table II**. Structural details: in-plane and out-of-plane lattice parameters of the $Sc_{1-x}V_xN_y$ and the $ScN_{0.87}$ samples with the corresponding volume. The average FWHM was obtained by fitting the φ-scans.

| Samples | Out-of-plane lattice parameter $a_\perp$ (±0.004) (Å) | In-plane lattice parameter $a_{//}$ (±0.004) (Å) | $a_\perp/a_{//}$ | Volume ($a_{//}^2 a_\perp$) (Å³) | Average FWHM in the φ-scans (°) |
|---|---|---|---|---|---|
| $ScN_{0.87}$ | 4.530 | 4.487 | 1.009 | 91.203 | 3.99 (±0.06) |
| $Sc_{0.96}V_{0.04}N_{0.85}$ | 4.522 | 4.480 | 1.009 | 90.758 | 4.82 (±0.09) |
| $Sc_{0.93}V_{0.07}N_{0.87}$ | 4.512 | 4.472 | 1.008 | 90.234 | 4.98 (±0.20) |
| $Sc_{0.91}V_{0.09}N_{0.83}$ | 4.502 | 4.461 | 1.009 | 89.592 | 5.35 (±0.13) |
| $Sc_{0.88}V_{0.12}N_{0.86}$ | 4.499 | 4.454 | 1.009 | 89.251 | 4.93 (±0.12) |
| $Sc_{0.85}V_{0.15}N_{0.81}$ | 4.484 | 4.445 | 1.008 | 88.594 | 5.03 (±0.18) |

## 3.4. Electronic structure of $Sc_{1-x}V_xN_y$

### 3.4.1. Polarization dependent X-ray Absorption Spectroscopy

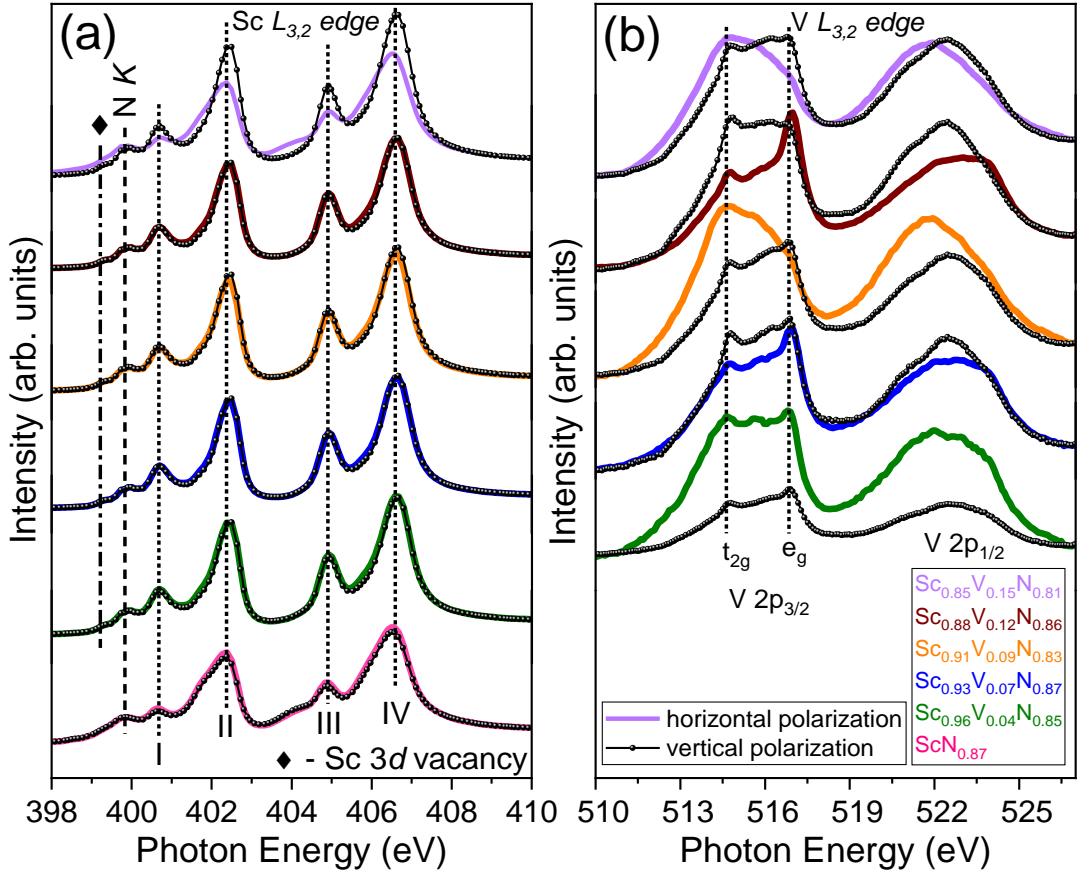

**Figure 3:** Polarization-dependent X-ray absorption measurements at the N *1s*-edge and the Sc *2p3/2,1/2*-edge (a) and the V *2p3/2,1/2*-edge (b) of the Sc$_{1-x}$V$_x$N$_y$ and the ScN$_{0.87}$ thin film samples. The data were normalized below and above each absorption edge.

Figure 3(a) and 3(b) shows polarization dependent normalized XAS data measured at the N *1s*-edge, Sc *2p3/2,1/2*-edge and V *2p3/2,1/2*-edge of Sc$_{1-x}$V$_x$N$_y$ and a reference ScN$_{0.87}$ sample. As observed in Figure 3(a), for all the samples, five prominent peaks appear at 399.8 eV, 400.6 eV (I), 402.3 eV (II), 404.9 eV (III) and 406.7 eV (IV). We assign the peak at 399.8 eV to the N *1s*-edge due to dipole allowed N *1s* → *2p* transitions ($\Delta l = \pm 1$) from the core to the valence states. The features I and II are attributed to Sc *2p3/2* - *t2g* and Sc *2p3/2* - *eg* states, whereas the higher energy features III and IV are attributed to *2p1/2* - *t2g* and *2p1/2* - *eg* states, respectively. The appearance of lower energy *t2g* states and higher energy *eg* states are due to the dominating octahedral symmetry with highly directional polar-covalent *eg* ($d_{z^2}$, $d_{x^2-y^2}$) orbitals strongly bonded to N $p_z$ orbitals forming σ-type bonds and metallic non-directional *t2g* ($d_{xy}$, $d_{xz}$, $d_{yz}$) orbitals forming π-type bonds with the N $p_x$ and the $p_y$ orbitals [59]. Due to the crystal field splitting (10*Dq*), both *2p3/2* and *2p1/2* peaks further split into *t2g* (Sc *2p3/2*→Sc *3d* +N *2pπ*) and *eg* (Sc *2p1/2*→Sc *3d* +N *2pσ*) peaks. The observed spin-orbit and crystal-field splitting are ~4.2 and ~1.7 eV, respectively [60,61]. The crystal field splitting directly scales with the bond length [62], and a larger crystal field splitting indicates stronger Sc *3d* - N *2p* bonding in the material. The extent of the crystal-field splitting depends on the coordination environment of the N ligands around the Sc ions in the crystal structure. The crystal field splitting [∝ (bond distance)$^{-5}$] appears to be rather constant within the energy resolution for the V alloying concentrations and does not increase with decrease in the lattice parameters.

In addition to a previous publication [60], we observe a distinct N *1s*-edge feature in the XAS spectra (see Figure 3(a)). The appearance of the distinct N *1s*-edge appears at the lower photon energy due to higher charge transfer to N ion from V and/or Sc ions compared to the absorption edge expected for a neutral N atom. It is noteworthy that in $Sc_{1-x}V_xN_y$ samples, an additional feature appears at around 399.2 eV. We attribute this feature to N vacancy and unintentional O induced localized Sc *3d* $t_{2g}^*$ defect states. The polarization dependent XAS studies at the N *1s* and the Sc *$2p_{3/2,1/2}$* edges indicate distinguishable intensity variations along horizontal and vertical polarizations only at the Sc *$2p_{3/2,1/2}$*-edge for $Sc_{0.85}V_{0.15}N_{0.81}$. The anisotropic contribution for the $Sc_{0.85}V_{0.15}N_{0.81}$ sample corresponds to antibonding π and σ-type orbitals due to distortion of the octahedral symmetry with different Sc-N bond distances in different crystallographic directions. The absence of intensity variation across the N *1s*-edge for the same sample is an indication of an isotropic distribution of N-N bonding.

The line shapes of the V *$2p_{3/2,1/2}$*-edge absorption in Figure 3(b) are less prominent than for a 100% occupancy of V at the metal site due to the low alloying concentration ($x$ = 0.04-0.15) in the $Sc_{1-x}V_xN_y$ samples. The doublet features around 515-516 eV, and 521-523 eV are attributed to the spin-orbit splitting of the *$2p_{3/2}$* and *$2p_{1/2}$* -edges with a *$t_{2g}$* - *$e_g$* sub splitting of ~2 eV. The polarization dependent XAS at V *$2p_{3/2,1/2}$*-edges also exhibits variable intensity distribution between the horizontal and the vertical polarizations. Unlike the Sc *$2p_{3/2,1/2}$*-edge, the anisotropic contribution is pronounced from the lowest up to the highest V alloying concentration for all the $Sc_{1-x}V_xN_y$ samples. This suggests that the V-V and V-N bonding is less uniformly ordered and more randomly distributed.

*3.4.2. Density of states calculations and Valence Band Spectroscopy*

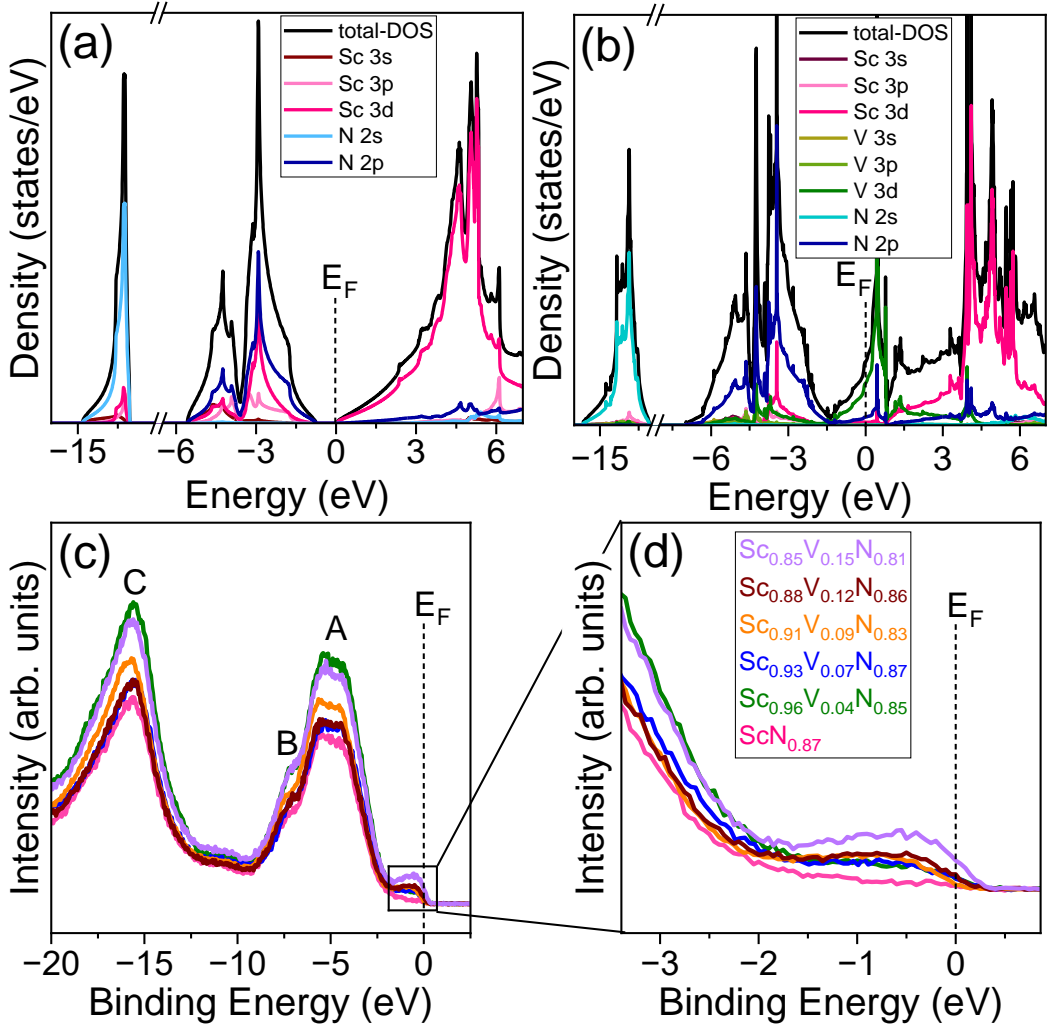

**Figure 4:** The total and partial-density of states (DOS) of (a) ScN and (b) $Sc_{0.75}V_{0.25}N$ calculated by DFT. (c) Valence band spectra of the $Sc_{1-x}V_xN_y$ and the reference $ScN_{0.87}$ thin film samples with different alloying concentrations. (d) A magnified view of the states near the Fermi level ($E_F$) in the VBS spectra measured by XPS.

Figure 4(a) and 4(b) shows the total and partial density of states (DOS) of $Sc_{0.75}V_{0.25}N$ in comparison to ScN. For comparison, experimental valence band spectra (VBS) of $Sc_{1-x}V_xN_y$ thin films and of a reference $ScN_{0.87}$ sample are shown in Fig. 4(c), while Fig. 4(d) shows a magnification of the total DOS for all the samples in the vicinity of the Fermi level ($E_F$). Due to the final state lifetime broadening and instrumental energy resolution, the experimental VBS occurs broader compared to the calculated DOS (see Fig. 4(a) and 4(b)). For the $ScN_{0.87}$ reference sample in Fig. 4(c), there are no states observed at the $E_F$ even for a sample containing N vacancies or O substitution. According to a previous report estimated from the theoretical calculations that combination of N vacancies and O substitution push the $E_F$ into the conduction band [20]. Thus, our observation suggests appearance of no additional defect states with the added electrons from O at $E_F$ around the valence band. An increasing V content in $Sc_{1-x}V_xN_y$ results in a gradual increase of intensity in the vicinity of the $E_F$ (see Fig. 4(c)). This can be understood as due to the contribution from two additional valence electrons from each V atom compared to ScN. The additional electrons from V drive the $Sc_{1-x}V_xN_y$ system towards a metallic-type of bonding. Since our samples contain finite amount of O, the contribution of O 2p states in the VBS spectra is discussed with the theoretical calculations in section 4 of the SI. However, no contribution from O appears at/near the $E_F$. Comparing the VBS data of $ScN_{0.87}$ with the DOS calculations, feature A and shoulder B are dominated by hybridization of N *2p* states with Sc *3d* states. Partial contribution from Sc *3p* states at feature A and Sc *3s* states at feature B can also be identified. The higher binding energy ($E_b$) feature C is mostly dominated by N *2s* states with partial contribution from Sc *3d* and *3p* states.

For the $Sc_{1-x}V_xN_y$ samples, feature A and B are due to hybridization of Sc *3d* and N *2p* states with partial contribution from V *3d* states. Minor contribution from V states can also be observed for the higher $E_b$ energy feature C apart from the main hybridizations which are similar to those of $ScN_{0.87}$. The shape of the VBS spectra remains similar while there is an increase in the intensity compared to $ScN_{0.87}$ that can be attributed to the presence of two additional valence electrons in V compared to Sc. For $Sc_{1-x}V_xN_y$, the calculated DOS across $E_F$ is dominated by V *3d* states with a low contribution from Sc *3d* and N *2p* states.

*3.4.5. Resonant Inelastic X-ray Scattering*

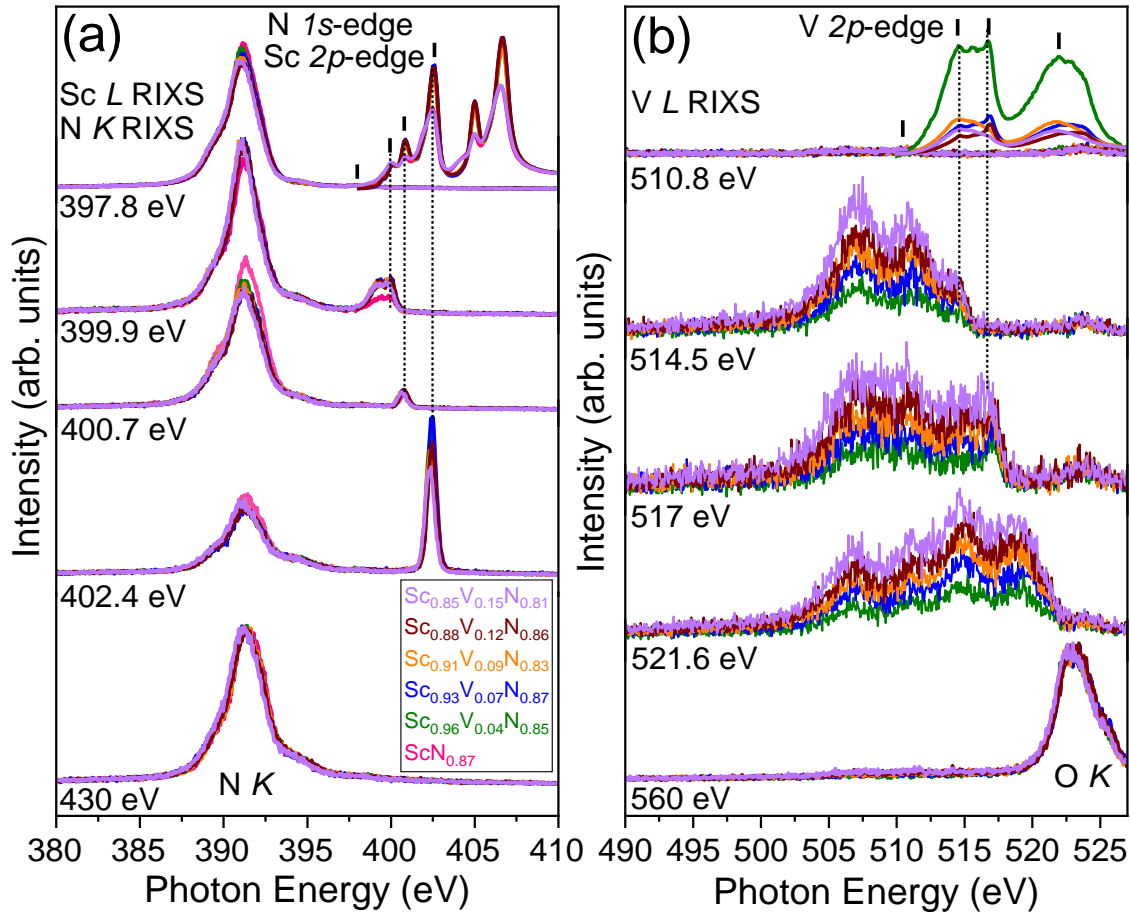

**Figure 5:** (a) Combined Sc *L* and N *K*-RIXS data for excitation energies 397.8 eV, 399.9 eV, 400.7 eV, 402.4 eV and 430 eV (non-resonant), respectively, and Sc *2p*-edge and N *1s*-edge XAS data (top right), (b) V *L* RIXS for excitation energies of 510.8 eV, 514.5 eV, 517 eV, 521.6 eV and 560 eV (non-resonant), respectively, and V *2p*-edge XAS (top right) of $Sc_{1-x}V_xN_y$ thin film samples compared to of $ScN_{0.87}$. The excitation energies for the normalized RIXS data are marked by the vertical bars in the XAS data.

Figure 5(a) shows Sc $L_{3,2}$ and N $K$-RIXS data of the Sc$_{1-x}$V$_x$N$_y$ thin film samples compared to ScN$_{0.87}$. The spectra were excited at 397.8 eV, 399.9 eV, 400.7 eV, 402.4 eV and 430 eV (non-resonant) photon energies, respectively. XAS data (top right) at the N $1s$-edge and Sc $2p$-edges were used to determine the absorption peak maxima used for the RIXS measurements.

Starting with the excitation energy of 397.8 eV, the main emission peak at 391.1 eV appears for all the samples due to N $2p$→$1s$ dipole transitions ($\Delta l = \pm 1$). Thus, the emission spectra are dominated by N $2p$ states, and the intensity variation depends on the difference in N content between the samples. Generally, the peak structures of the ScN$_{0.87}$ reference sample are consistent with a previous publication, where the main peak is accompanied by a shoulder peak at ~389.2 eV [61] and the same trend is observed for the Sc$_{1-x}$V$_x$N$_y$ samples. This is attributed to the strong Sc $3d$ – N $2p$ hybridization. A weak emission feature at around 394.3 eV occurs due to the Sc $L_3$ (Sc $3d4s$→$2p_{3/2}$) emission line.

As the photon energy is increased to 399.9 eV corresponding to N $1s$→$2p$ absorption in the XAS, a doublet feature appears at around 399-400 eV in addition to the main emission peak at 391.1 eV and Sc $L_3$ emission. In the doublet, the highest energy peak at 399.9 eV corresponds to elastic Rayleigh scattering and the lower energy peak at 399.2 eV is due to resonance with the N $1s$ XAS resulting in the appearance of N $2p$ states hybridized with Sc $3d$ states. As the photon energy is further raised to 400.7 eV, corresponding to the Sc $2p_{3/2}$-$t_{2g}$ peak in XAS, the elastic Rayleigh scattering peak appears at 400.7 eV in addition to the main emission peak at 391.1 eV and Sc $L_3$ emission. As the photon energy is further raised to 402.4 eV, the elastic peak becomes significantly stronger. Strong Rayleigh scattering is a signature of localized electron excitations in narrow band systems that is normally observed in highly correlated rare earth systems [63] and corroborates to the covalent bonding for $L_3$-$e_g^*$ states. For comparison, the relatively weak elastic peaks corresponding to $L_3$-$t_{2g}^*$ signifies delocalization that should affect the electrical conductivity. At non-resonant emission of 430 eV excitation energy, only the N $2p$ and Sc $L_3$ emission lines are observed with no additional features.

Figure 5(b) shows V $L$ RIXS for excitation energies of 510.8 eV, 514.5 eV, 517.0 eV, 521.6 eV and 560.0 eV (non-resonant), respectively, at the V $2p$ absorption edge (top right). For the pre-peak excitation, the emission is very weak while at 514.5 eV corresponding to the $2p_{3/2}$-$t_{2g}$ excitation energy, three broad features are observed. The highest energy feature at 514.5 eV corresponds to elastic Rayleigh scattering while the strong doublet is due to $L_3$-$t_{2g}$ and $L_3$-$e_g$ having a crystal field splitting of 3.7 eV. As the energy is raised to 517.0 eV, corresponding to the $2p_{3/2}$-$e_g$ excitation energy, the double peak is less pronounced. As the excitation energy is raised to 521.6 eV corresponding to the V $2p_{1/2}$ shell, four distinct features appear where the lower energy doublet corresponds to $L_3$-$t_{2g}$ and $L_3$-$e_g$ and the higher energy doublet are $L_2$-$t_{2g}$ and $L_2$-$e_g$ due to V $3d4s$→$2p_{3/2,p1/2}$ dipole allowed transitions. Moreover, due to excitation at the V $L_2$ absorption edge, the V $L_2$ features are more intense compared to the V $L_3$ features. The gradual increase in intensity from Sc$_{0.96}$V$_{0.04}$N$_{0.85}$ to Sc$_{0.85}$V$_{0.15}$N$_{0.81}$ at 514.5 eV, 517 eV, and at 521.6 eV excitation energies is associated with the increasing alloying concentration of V. Finally, at non-resonant excitation at 560.0 eV, the spectra become dominated by the O $K$ emission at 523 eV. However, weak O $K$ emission is present in all three 514.5, 517 and 521.6 eV emission energies. This further depicts that our samples have a finite amount of oxygen as was confirmed from the ERDA measurements mentioned in section 3.2.

## 3.4. Electrical properties

**Table II**: Seebeck coefficient (S), electrical resistivity (ρ), power factor ($S^2\sigma$), charge carrier concentration (n) and charge carrier mobility (μ) of the reference $ScN_{0.87}$ and $Sc_{1-x}V_xN_y$ samples at room temperature.

| Samples | S (μV K$^{-1}$) | ρ (μΩcm) | $S^2\sigma$ (μW cm$^{-1}$ K$^{-2}$) | n (cm$^{-3}$) | μ (cm$^2$ V$^{-1}$ s$^{-1}$) |
|---|---|---|---|---|---|
| $ScN_{0.87}$ | -21 ± 7 | 140 | 3.15 | $-2.7 \times 10^{21}$ | 16.51 |
| $Sc_{0.96}V_{0.04}N_{0.85}$ | -7 ± 6 | 1040 | 0.047 | $-2.85 \times 10^{21}$ | 2.11 |
| $Sc_{0.93}V_{0.07}N_{0.87}$ | -4 ± 7 | 1520 | 0.011 | $-3.86 \times 10^{21}$ | 1.1 |
| $Sc_{0.91}V_{0.09}N_{0.83}$ | -5 ± 7 | 2670 | 0.009 | $-3.67 \times 10^{21}$ | 0.63 |
| $Sc_{0.88}V_{0.12}N_{0.86}$ | -1 ± 4 | 2270 | 0.0004 | $-3.53 \times 10^{21}$ | 0.78 |
| $Sc_{0.85}V_{0.15}N_{0.81}$ | 10 ± 4 | 3010 | 0.033 | $8.83 \times 10^{21}$ | 2.35 |

Electrical properties (Seebeck coefficient, electrical resistivity, power factor, charge carrier concentration and carrier mobility) studied at room temperature are tabulated in Table II. The reference $ScN_{0.87}$ shows electron-dominated charge transport with $S \sim -21$ μV K$^{-1}$ which is in close agreement with available reports [64]. With V increase in $Sc_{1-x}V_xN_y$ up to $x = 0.12$, Seebeck coefficient remains negative with their absolute values decreasing towards zero. For the highest V containing $Sc_{1-x}V_xN_y$ with $x = 0.15$, the measured Seebeck coefficient changes to positive with a value around +10 μV K$^{-1}$. The electrical resistivity values of $Sc_{1-x}V_xN_y$ samples increases gradually from 1040 to 3010 μΩcm with increase in alloying concentration from $x = 0.04$ to 0.12, an order of magnitude higher than 140 μΩcm for the reference $ScN_{0.87}$ sample. The absolute values of the charge carrier concentration remain nearly the same at $(2.7 - 2.85) \times 10^{21}$ cm$^{-3}$ for reference $ScN_{0.87}$ and $Sc_{1-x}V_xN_y$ with $x = 0.04$. A slight increase in the carrier concentration is observed with higher V content and the values nearly remain the same within $(3.53 - 3.86) \times 10^{21}$ cm$^{-3}$ for $Sc_{1-x}V_xN_y$ with $x = 0.07 - 0.12$. For $Sc_{1-x}V_xN_y$ up to $x = 0.12$ including the reference $ScN_{0.87}$, the carrier concentration value is negative. For the highest V content ($x = 0.15$), there is at least two times increase in the carrier concentration with a positive value of $8.83 \times 10^{21}$ cm$^{-3}$. Simultaneously, there is an abrupt decrease of the charge carrier mobility (2.35 - 0.78 cm$^2$ V$^{-1}$ s$^{-1}$) for $Sc_{1-x}V_xN_y$ compared to reference $ScN_{0.87}$ (16.51 cm$^2$ V$^{-1}$ s$^{-1}$). The increase in $\rho$ with V content might be due to increased local disorder which is reflected as reduction in carrier mobilities as well. The combined effect of reduced $S$ and increased $\rho$, results in an overall decrease in $S^2\sigma$ from 3.15 μW cm$^{-1}$ K$^{-2}$ for reference $ScN_{0.87}$ to a range of 0.047-0.0004 μW cm$^{-1}$ K$^{-2}$ for $Sc_{1-x}V_xN_y$ samples.

## 4. Discussion

The GGA+U (U = 3.5 eV) calculations reveal that ScN exhibits a relaxed lattice parameter of 4.518 Å with cubic B1 NaCl structure (see Figure 1(c)). Substitution of 1 Sc atom by V in the unit cell of ScN (*i.e.*, 3 Sc atoms, 1 V atom and 4 N atoms) still retains the rocksalt B1 structure (see Figure 1(d)) with a reduction in the lattice parameter to 4.410 Å in relaxed state. Due to two additional valence electrons of V compared to Sc, the electron concentration of the $Sc_{0.75}V_{0.25}N$ system is expected to be higher, contributing higher DOS across the $E_F$ and thus higher electrical conductivity compared to ScN.

For the $Sc_{0.75}V_{0.25}N$ system (Fig. 1(b)), the $E_F$ shifts into the CB and an overlap of the CB and VB occurs at Γ-point. This is different from the direct (Γ-Γ and X-X symmetry points) and the indirect (Γ-X) bandgaps observed for ScN (see Figure 1(a)). The main contribution in the vicinity of the $E_F$ originates from V $3d_{xy}$, $3d_{xz}$, $3d_{yz}$ bands crossing the $E_F$ halfway at the Γ-point. The V $3d$ states are dispersive at the Γ-point with band multiplicity indicating towards achieving higher σ and a simultaneous occurrence of flat bands along the X-W-K direction insights for inducing higher $S$ compared to ScN. Thus, the $Sc_{1-x}V_xN$ system shows promise as an efficient thermoelectric material compared to ScN. However, we limited the alloying concentration of V to up to 15% in experiments (and not 25% as calculated theoretically). The reason is to take into consideration the effects of induced defects during the non-equilibrium sputtering process which could not be neglected. The N vacancies and incorporation of unintentional O are especially thermodynamically favorable in the ScN system [20,65]. Thus, the increase in induced defects with increasing alloying concentration in principle could lead to an adverse effect on the thermoelectric properties.

This is supported by a series of $Sc_{1-x}V_xN_y$ (x = 0.04-0.15) thin film samples and reference ScN film with N vacancies deposited on MgO(001) substrates. Compositional analysis showed that all the samples were N deficient (y = 0.81-0.87) with presence of unintentional O (5 to 8.2 at. %) across the N-site. The origin of O incorporation could be the adaptation of high substrate temperature leading to higher amount of degassing within the chamber and/or air exposure post deposition. It is to be noted that the waiting time to minimize the degassing effect was around 30 mins with an achieved base pressure of ≤1.3×10$^{-6}$ Pa (9.6×10$^{-9}$ Torr) prior to the depositions. From the XRD measurements, no additional phases were observed for $Sc_{1-x}V_xN_y$ samples exhibiting epitaxial nature as observed from the φ-scans (see Fig. 2(c)). The in-plane and out-of-plane XRD revealed that all the samples exhibited slight distortion in the cubic structure with an elongation along the *c*-axis.

The XAS data reveals presence of localized Sc $3d$ $t_{2g}$ states in $Sc_{1-x}V_xN_y$ induced due to N vacancies and presence of unintentional O which is the most stable defect configuration in ScN system [66]. However, the defect states in reference $ScN_{0.87}$ cannot be separated out from the N *K*-edge. The presence of N vacancies results in excess valence electrons along the metal site (Sc+V) which remains unbounded. However, presence of more electronegative O (along with N at the non-metal site) than N can accommodate two valence electrons/atom but O itself also has an additional valence electron than N. Therefore, an excess of electrons should result in lifting the Fermi level to the conduction band as indicated by theoretical band structure calculations [20]. In comparison to a previous study [60], we found a distinct N *1s*-edge for the reference $ScN_{0.87}$ sample in XAS (Fig. 3(a)) and similar observations were also found for the $Sc_{1-x}V_xN_y$ samples. The appearance of a distinct N *1s*-edge is expected at the lower photon energy due to higher charge transfer to the N ions from V and/or Sc ions compared to the absorption edge expected for neutral N atoms.

Note that our XRD results show pseudocubic crystal structures for all the samples but polarization dependent XAS reveals the independent contribution of the bonding anisotropy. The polarization dependent XAS study revealed a small distortion of the octahedral symmetry for the highest alloying $Sc_{0.85}V_{0.15}N_{0.81}$ sample at the Sc $2p_{3/2,1/2}$-edge due to variable intensity distribution in horizontal and vertical polarization (Fig. 3(a)). However, at the V $2p_{3/2,1/2}$-edge, distinguishable intensity distribution in both the polarization states was found for the lowest V alloying concentration (see figure 3(b)). These observations signify different Sc-N bond lengths for $Sc_{0.85}V_{0.15}N_{0.81}$ only, while all the V alloyed samples had variable V-V and V-N bond distances along the x, y, and z directions. On the contrary, the distribution of N-N bonds appeared to be isotropic. The difference between the two polarizations (vertical and horizontal) indicates that there may be not only N *2p* states but also some contribution of O *2p* states bonding to the *3d* states of the Sc and V atoms as ligands. This results in slight local distortion of the octahedral symmetry. Thus, we infer that the distortion in the cubic structure of the Sc$_{1-x}$V$_x$N$_y$ samples is primarily due to anisotropic V-N and V-V bonds. However, for the highest V alloying concentration, the anisotropic Sc-N bonds also contribute to the distortion in the octahedral symmetry.

The combined VBS results (Fig. 4) and DOS calculations show that all the Sc$_{1-x}$V$_x$N$_y$ samples have finite density of states across the $E_F$. This population is dominated by V *3d* states with a minor contribution from Sc *3d* and N and O *2p* states. No states are observed for reference ScN$_{0.87}$ from VBS. This is noteworthy, because one might expect features near $E_F$ and indicates that the added electrons from O do not add features of a finite DOS at $E_F$. Nevertheless, it has been indicated in literature that Sc vacancies accompanied with substitutional oxygen is a stable defect [66]. These defects push $E_F$ inside the conduction band resulting in finite DOS contribution [20]. The appearance of weak elastic Rayleigh scattering peaks (Fig. 5a) corresponding to $L_3$-$t_{2g}^*$ at resonant photon energies of 399.9 eV and 400.7 eV are noted. This signifies delocalization of the Sc$_{1-x}$V$_x$N$_y$ and reference ScN$_{0.87}$ systems, indicating contribution of free mobile charge carriers for the electrical conductivity. Indeed, at 399.9 eV, the intensity of ScN$_{0.87}$ is lower compared to Sc$_{1-x}$V$_x$N$_y$ samples signifying higher conductivity of ScN$_{0.87}$ compared to Sc$_{1-x}$V$_x$N$_y$. In contrast, at 402.4 eV resonant energy, the strong Rayleigh scattering is a signature of localized electron excitations corroborating to the covalent bonding for $L_3$-$e_g^*$ states. The covalent bonding weakens with higher V alloying. The increase of intensity of the V *2p* emission lines at 514.5, 517 and 521.6 eV excitation energies with higher V alloying (Fig. 5b) further confirms consistent results with the EDS and the ERDA results. The presence of O is also observed at non-resonant 560 eV photon energy in-line with our compositional analysis.

The transport properties of ScN are highly susceptible to the presence of point defects such as vacancies (Sc and N), interstitials (N) and substitutional defects (O). The electrical resistivity and Seebeck coefficient (hold for metals and degenerate semiconductors) are related to the carrier concentration (here for electrons *n* but the relations hold for holes *p* as well) as [3,67],

$$\rho = \frac{1}{ne\mu}$$

and,

$$S = \frac{8\pi^2 k_B^2 T}{3eh^2} m^* \left(\frac{\pi}{3n}\right)^{2/3}$$

where, the contribution from majority (electrons/holes) and minority (holes/electrons) charge carriers are [67],

$$S = \frac{S_n \sigma_n + S_p \sigma_p}{\sigma_n + \sigma_p}$$

and the subscripts represent the type of charge carriers (n: electrons, p: holes). Here, $k_B$ = Boltzmann's constant, $e$ = electronic charge, $h$ = Planck's constant, and $m^*$ = effective mass of the charge carrier at an absolute temperature *T*. Based on the Mahan-Sofo theory, the *DOS* is associated with the effective mass as [2,68],

$$DOS(E) = \frac{(m_d^*)^{3/2}\sqrt{2E}}{\hbar^3 \pi^2}$$

Further, it is to be noted here that although charge carriers exhibiting higher effective mass will have lower mobility but it also depends on several other factors like scattering, electronic structure and anisotropy behavior of the crystal system [3].

For $Sc_{1-x}V_xN_y$ samples, up to $x = 0.12$, a reduction is observed in the absolute $S$ values compared to $ScN_{0.87}$ by an order of magnitude, the values typical for metals [69]. It is worth noting that the charge carrier concentration remains relatively constant for $x = 0.04$ and reference $ScN_{0.87}$ to $x = 0.07 - 0.12$ to $x = 0.15$ suggests a dominant role of the charge carrier mobility to their contribution in the Seebeck values. Although the $S$ values for $Sc_{0.91}V_{0.09}N_{0.83}$ and $Sc_{0.88}V_{0.12}N_{0.86}$ samples are at the borderline between positive and negative considering the error bars, the sign of the carrier concentration confirms that the majority charge carriers are electrons for these samples. Later interestingly the charge transport becomes hole-dominated for the highest V content ($Sc_{0.85}V_{0.15}N_{0.81}$). The observed acceptor nature can be attributed to possible Sc vacancies depending on the defect chemistry of ScN having Sc vacancies of acceptor nature while N vacancies and interstitials are triple and single donors, respectively [65].

The obtained charge carrier concentration of $\sim 10^{21}$ cm$^{-3}$ is typical for metals and heavily doped semiconductors [3]. Since in our VBS results we observed an increase in DOS across $E_F$ for $Sc_{1-x}V_xN_y$ samples compared to $ScN_{0.87}$, it suggests a heavier effective mass of electrons in V alloying samples. This reflects in decrease of carrier mobility from an order to even two orders of magnitude lower compared to $ScN_{0.87}$. This could be attributed to the presence of defect concentrations such as Sc and N vacancies, V alloying and unintentional O across the non-metal site. Each of them behaves as scattering centers for the electrons which reflects not only in the lowering of the charge carrier mobility but also to an increase in the overall resistivity of the samples by at least an order of magnitude higher compared to $ScN_{0.87}$. Consequently, a reduction in the thermoelectric power factor has been observed.

For the highest alloying sample $Sc_{0.85}V_{0.15}N_{0.81}$, holes instead contribute as the majority charge carriers, as indicated by the positive Seebeck coefficient in this sample. From the Hall measurements, the carrier concentration was found to be at least two times higher, and the mobility increased by an order of magnitude compared to the second highest alloying sample $Sc_{0.88}V_{0.12}N_{0.86}$. It was also reflected in an increase of the electrical resistivity for this sample. Essentially the $n$ to $p$ type switching here is driven by lower concentration of less mobile electrons which stems from the highest N and Sc vacancies incorporated by O and V along the non-metal and metal site.

## 5. Conclusions

In conclusion, the electronic structure of epitaxial $Sc_{1-x}V_xN_y$ films with different alloying concentrations of V were investigated. All samples exhibit N substoichiometry with unintentional presence of O along the non-metal site suggesting N substitution. The lattice parameter gradually reduces compared to reference $ScN_{0.87}$ obeying Vegard's law. Up to the highest alloying concentration, the $Sc_{1-x}V_xN_y$ samples retain a NaCl structure. However, the N-N bond length is symmetrically distributed in the 3D co-ordinate but an asymmetry in the Sc-N bond length is observed for the highest V content ($x = 0.15$). On the contrary, the V-V and V-N bond lengths vary randomly in all the three-co-ordinate axis. Moreover, N induced Sc $3d$ $t_{2g}^*$ states were observed for the first time experimentally from the Sc *2p* XAS, in line with the earlier theoretical reports. Band structure calculations demonstrate that for Sc$_{0.75}$V$_{0.25}$N an overlap of the conduction and valence band occurs, and the Fermi level lies in the conduction band. Thus, the system is driven to metal-like behavior from semiconducting ScN. The presence of light (dispersive) bands at Γ-point with band multiplicity leads to higher electrical conductivity and flat (heavy) bands at X-W-K symmetry points signify higher Seebeck coefficient than ScN demonstrating the potential as a thermoelectric material. Moreover, finite density of states crossing the Fermi level in Sc$_{1-x}$V$_x$N$_y$ is due to contribution from V *3d* states. However, experimentally we observed an increase in the electrical resistivity accompanied with a decrease in the negative Seebeck coefficient. The Hall measurements confirm the reduced charge carrier mobility due to higher scattering by defects to be the primary reason for such a decrease in the thermoelectric power factor. In addition, the mobility reduces so much that instead of the electrons, holes become the majority charge carriers for the highest V alloying sample.

## Acknowledgements

The authors acknowledge funding from the Swedish Government Strategic Research Area in Materials Science on Functional Materials at Linköping University (Faculty Grant SFO-Mat-LiU No. 2009 00971), the Knut and Alice Wallenberg foundation through the Wallenberg Academy Fellows program (KAW-2020.0196), the Swedish Research Council (VR) under Project No. 2021-03826. SC acknowledges Victor Hjort for his help during depositions. M.M. also acknowledges financial support from the Swedish Energy Agency (Grant No. 43606-1) and the Carl Tryggers Foundation (CTS23:2746, CTS20:272, CTS16:303, CTS14:310). GG acknowledges the Swedish Energy Agency project 51201-1, the Åforsk Foundation Grant 22-4, and the Olle Enqvist foundation grant 222-0053.

Research conducted at MAX IV, a Swedish national user facility, is supported by the Swedish Research council under contract 2018-07152, the Swedish Governmental Agency for Innovation Systems under contract 2018-04969, and Formas under contract 2019-02496. Anirudha Ghosh and Conny Såthe from SPECIES beamline, MAX IV, Lund, Sweden is acknowledged for the help during the XAS and RIXS measurements. The FEFF calculations were enabled by resources provided by the National Academic Infrastructure for Supercomputing in Sweden (NAISS) at Linköping University, partially funded by the Swedish Research Council through grant agreement no. 2022-06725. Daniel Primetzhofer from Uppsala University is acknowledged for Accelerator operation supported by Swedish Research Council VR-RFI (Contract No. 2019-00191) and the Swedish Foundation for Strategic Research (Contract No. RIF14-0053).

Supplementary Information

# Electronic structure and thermoelectric properties of epitaxial Sc$_{1-x}$V$_x$N$_y$ thin films grown on MgO(001)

Susmita Chowdhury[*], Niraj Kumar Singh, Sanath Kumar Honnali, Grzegorz Greczynski, Per Eklund, Arnaud le Febvrier, and Martin Magnuson

*Thin Film Physics Division, Department of Physics, Chemistry and Biology (IFM), Linköping University, Linköping SE-581 83, Sweden*

## 1. Rutherford Backscattering Spectrometry (RBS)

*The ERDA results indicate qualitatively about the presence of a heavier element than Sc and V in Sc$_{1-x}$V$_x$N$_y$ samples and reference ScN$_{0.87}$. Therefore,* Rutherford backscattering spectrometry (RBS) measurements were performed *on all the samples* to quantify the trace of heavy element contamination observed in ERDA. The RBS measurements were conducted with 2 MeV He$^+$ primary ions incident at 5° to the sample surface normal and the resulting spectra was analyzed using the *SIMNRA* 7.03 software [1]. *From fitting of the experimental data (not shown here), a trace amount of Hf was found to be present in all the samples with ≈0.1 atomic %.*

## 2. X-ray Reflectivity (XRR)

Figure S1 shows the X-ray reflectivity (XRR) of reference ScN$_{0.87}$ and Sc$_{1-x}$V$_x$N$_y$ thin film samples deposited on MgO(001) substrates. The measurements were performed using PANalytical X'Pert Pro diffractometer equipped with Cu-Kα source. The fitting of the XRR spectra were performed using Parratt 32 software based on Parrat's formalism [2] to obtain the thickness and real part of the scattering length density (real SLD) of the samples. The real SLD (α) is related to the critical momentum transfer vector q$_c$ = 4πsinθ$_c$/λ where, θ$_c$ = (λ$^2$α/π)$^{1/2}$. Here, θ$_c$ = critical angle, λ = wavelength of X-rays. α is related to the mass density (d) as, d = αM/N$_A$r$_e$Z where, M is molar mass, N$_A$ is Avogadro's number (6.023×10$^{23}$ mol$^{-1}$), Z is atomic number and r$_e$ is classical electron radius (2.818 fm) [3]. The fitting parameters are tabulated in Table S1. The sample thickness of reference ScN$_{0.87}$ is ~129 nm while it varies between ~125 to 130 nm for Sc$_{1-x}$V$_x$N$_y$ samples. With V alloying, there is a gradual increase in real SLD which is expected as V has higher mass density than Sc.

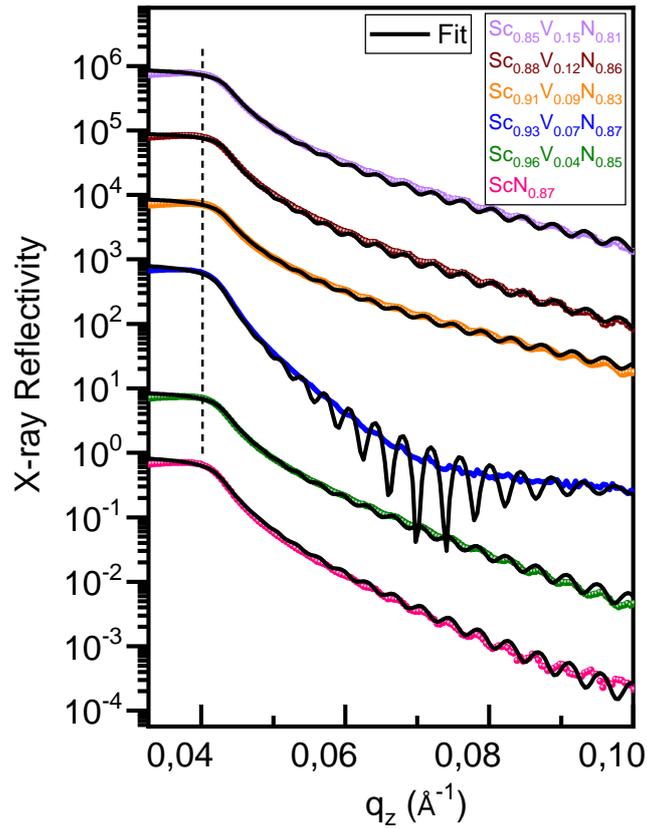

**Figure S1.** XRR spectra of reference $ScN_{0.87}$ and $Sc_{1-x}V_xN_y$ thin film samples with different alloying concentrations.

**Table S1**. The thickness and real part of the scattering length density (SLD) of the thin film samples obtained from fitting of the XRR spectra.

| Samples | XRR analysis | |
|---|---|---|
| | Thickness (nm) | Real SLD ($\pm0.02\times10^{-5}$) (Å$^{-2}$) |
| $ScN_{0.87}$ | 129 ± 2 | 3.62×10$^{-5}$ |
| $Sc_{0.96}V_{0.04}N_{0.85}$ | 130 ± 2 | 3.70×10$^{-5}$ |
| $Sc_{0.93}V_{0.07}N_{0.87}$ | 129 ± 6 | 3.70×10$^{-5}$ |
| $Sc_{0.91}V_{0.09}N_{0.83}$ | 129 ± 2 | 3.72×10$^{-5}$ |
| $Sc_{0.88}V_{0.12}N_{0.86}$ | 125 ± 2 | 3.74×10$^{-5}$ |
| $Sc_{0.85}V_{0.15}N_{0.81}$ | 126 ± 2 | 3.78×10$^{-5}$ |

## 3. Core level X-ray Photoelectron Spectroscopy (XPS)

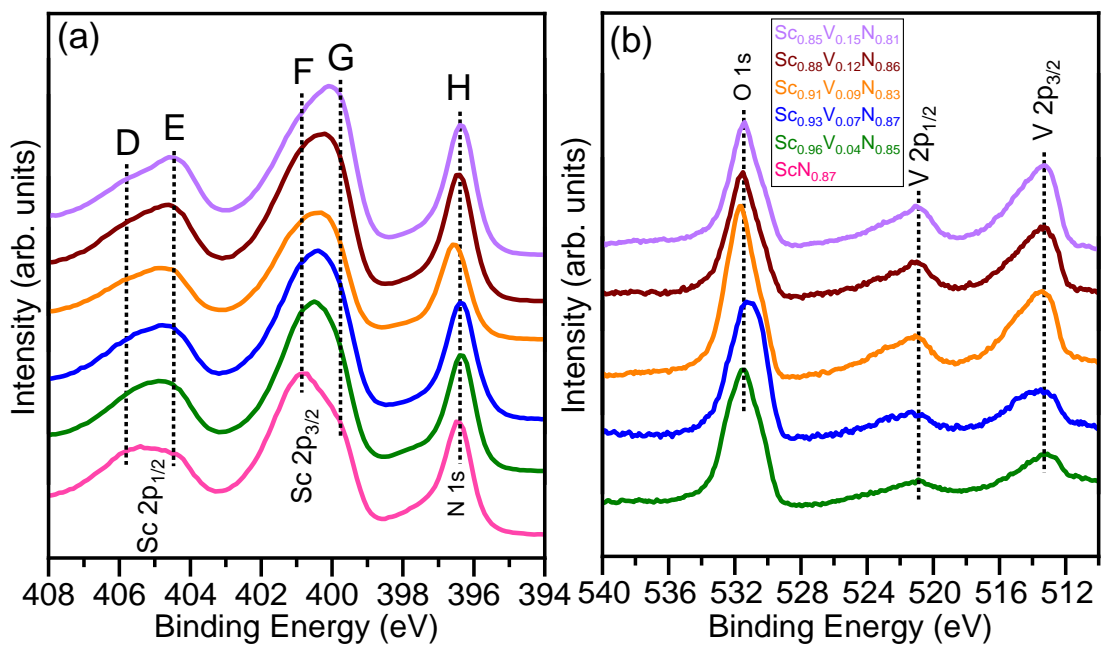

**Figure S2.** XPS core level spectra of Sc *2p* and N 1s (a) and V *2p* (b) of reference ScN$_{0.87}$ and Sc$_{1-x}$V$_x$N$_y$ thin film samples with different alloying concentrations.

Figure S2 shows the Sc *2p*, N *1s* and V *2p*, O 1s XPS core level spectra normalized to the highest intensity. For the reference ScN$_{0.87}$ sample (see Figure S2(a)), the N *1s* peaks at around 396.4 eV and due to spin split doublet peaks Sc *2p$_{3/2}$* (~400.8 eV) and Sc *2p$_{1/2}$* (~405.2 eV) appears for Sc. The Sc *2p$_{1/2}$* is more broadened compared to Sc *2p$_{3/2}$* due to the Coster-Kroning effect [4]. The spin orbit splitting of ~4.4 eV is same as obtained in XAS within the instrumental resolution. An asymmetry at lower BE side is observed around both Sc *2p$_{3/2}$* and Sc *2p$_{1/2}$*. This can be attributed to the finite presence of Sc-O-N$_x$/Sc-O in addition to Sc-N bonding. For Sc$_{1-x}$V$_x$N$_y$, the contribution of Sc-O-N$_x$/Sc-O bonding is more pronounced around the Sc *2p$_{3/2}$* and Sc *2p$_{1/2}$* spectra. An interplay between the intensities around D, E, F and G is noted. Initially for reference ScN$_{0.87}$, the features D and F were intense. However, with increase in alloying concentration of V, the features E and G become more intense. No visible shift around N *1s* is observed for Sc$_{1-x}$V$_x$N$_y$ samples compared to reference ScN$_{0.87}$ except the sample with $x = 0.09$. An asymmetry noted in all the samples around the lower BE of N *1s* is assigned to the presence of ScO$_x$N$_y$ bonds [5].

Figure S2(b) shows doublet V *2p$_{3/2}$* and V *2p$_{1/2}$* due to spin split (~7.6 eV) in Sc$_{1-x}$V$_x$N$_y$ samples. The broadening and asymmetry in V *2p* core level spectrum is resultant of the multiplet effect due to unpaired electrons in V$^{\delta+}$ state [6,7]. No significant shift in the V *2p* spectra can be noted for all the Sc$_{1-x}$V$_x$N$_y$ samples. The O *1s* core level spectra shows no visible peak shift within Sc$_{1-x}$V$_x$N$_y$ samples and reference ScN$_{0.87}$ suggesting no change in the oxidation state of O.

## 4. Theoretical DOS with N vacancy and O

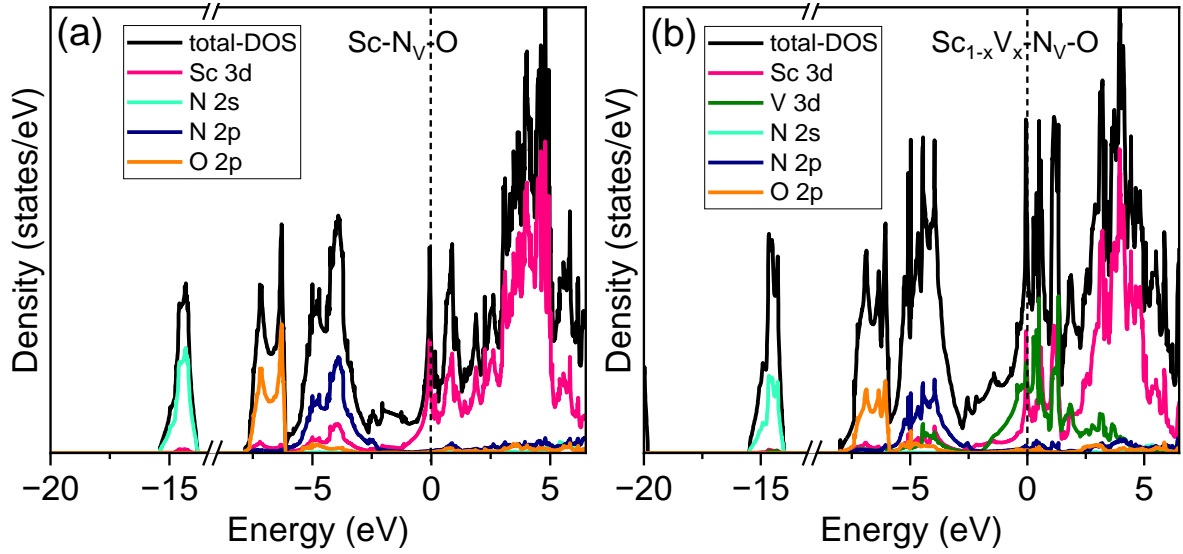

**Figure S3.** Theoretical density of states (DOS) considering N vacancies and O in the unit cell of ScN and $Sc_{1-x}V_xN$.

Figure S3 shows the theoretical DOS including N vacancies and O in the unit cell of ScN (see Fig. S3(a)) and $Sc_{1-x}V_xN$ (see Fig. S3(b)). As can be seen, N vacancies induce a finite Sc 3d $t_{2g}^*$ states across the Fermi level. The observation was noted in the unoccupied Sc *2p* XAS spectra mentioned in the main manuscript. We do not observe any significant contribution from O 2p states at or around the Fermi level. The contribution can only be noted at around 6-8 eV in the DOS calculations. Therefore, we attribute a contribution of O 2p states around feature B in Figure 4(c) mentioned in the main manuscript.